\documentclass[aps,prb,floats,showpacs,twocolumn]{revtex4}
\usepackage{graphicx,soul,color,mathrsfs,wrapfig,amsmath}

\ifx\pdftexversion\undefined
\usepackage[dvips]{hyperref}
\else
\usepackage{hyperref}
\fi
\hypersetup{
  colorlinks = true, linkcolor = magenta
}
\def \tr{{\mbox{tr~}}}
\def \ra{{\rightarrow}}
\def \ua{{\uparrow}}
\def \da{{\downarrow}}
\def \be{\begin{equation}}
\def \ee{\end{equation}}
\def \ba{\begin{array}}
\def \ea{\end{array}}
\def \bea{\begin{eqnarray}}
\def \eea{\end{eqnarray}}
\def \nn{\nonumber}

\def \e{{\epsilon}}

\def \a{{\alpha}}

\def \b{{\beta}}
\def \g{{\gamma}}
\def \D{{\Delta}}
\def \d{{\delta}}

\def \s{{\sigma}}

\def \e{{\epsilon}}

\def \G{{\Gamma}}

\def \nd{{^{\vphantom{\dagger}}}}
\def \yd{^\dagger}

\begin{document}

%%%%%%%%%%%%%%%%%%%%%%%
\title{Enhancement of the critical temperature in cuprate superconductors by inhomogeneous doping}
\author{Lilach Goren}
\affiliation{Department of Condensed Matter Physics, The Weizmann Institute of Science, 76100 Rehovot (Israel)}
\author{Ehud Altman}
\affiliation{Department of Condensed Matter Physics, The Weizmann Institute of Science, 76100 Rehovot (Israel)}
\date{\today}
\begin{abstract}
%%%%%%%%%%%%%%%%%%%%%%%
We use a renormalized mean field theory to investigate the superconducting properties of underdoped cuprates embedded with overdoped or metallic regions that carry excess dopants.
The overdoped regions are considered, within two different models, first as stripes of mesoscopic size larger than the coherence length and then as point impurities. In the former case we compute the temperature dependent superfluid stiffness by solving Bogoliubov de Gennes equations within the slave boson mean field theory. We average over stripes of different orientations to obtain an isotropic result. To compute the superfluid stiffness in the model with point impurities we resort to a diagrammatic expansion in the impurity concentration (to first order) and their strength (up to second order). We find analytic expressions for the disorder averaged superfluid stiffness and the critical temperature. For both types of inhomogeneity we find increased superfluid stiffness, and for a wide range of doping enhancement of $T_c$ relative to a homogeneously underdoped system. Remarkably,  in the case of microscopic impurities we find that the maximal $T_c$ can be significantly increased compared to $T_c$ at optimal doping of a pure system.
\end{abstract}
\maketitle
%%%%%%%%%%%%%%%%%%%%%%%%%%%%%%%%%%%%%%%%%%%%%%
\section{Introduction}\label{section:introduction}

Local probes of the cuprate superconductors reveal signatures of
electronic inhomogeneity both at the microscopic scales of lattice constants and at somewhat larger mesoscopic scales \cite{Chang:1992,Pan:2001,Howald:2001,Gomes:2007,Koshaka:2007,Pasupathy:2008,Parker:2010}.
The inhomogeneity is generally seen as separated regions with either a large or a small gap, which
have been attributed to local variations in the doping level with respect to the half filled Mott insulator\cite{Parker:2010}.

Experiments which probe global properties indicate that the average doping level has two direct effects on the superconducting properties. First, the pairing gap is seen to decrease with hole doping away from half filling\cite{Ding:2001,Ino:2002}. Second, the superfluid stiffness extracted from penetration depth measurements, increases with doping\cite{Uemura:1989,Boyce:2000}. This interplay between two energy scales relevant to superconductivity is thought to give rise to the dome shaped dependence of $T_c$ on hole doping\cite{Emery:1995}. Doping inhomogeneity is therefore expected to  lead to spatial modulations of the pairing amplitude along with variations of the charge carrier density.

In this paper we shall investigate how inhomogeneity in the doping level affects global superconducting properties of the material. Specifically we address the effect of inhomogeneity on the temperature dependent thermodynamic stiffness and, ultimately, on the transition temperature. To this end we employ a semi-phenomenological model of a $d$-wave superconductor that takes into account the
the proximity to the Mott insulator through a strong on-site repulsion. Furthermore we consider
various scales of inhomogeneities, ranging from the microscopic scale of a lattice constant to mesoscopic scales, somewhat larger than the coherence length (see Fig \ref{fig:overview}). An important question for practical applications is whether the transition temperature can be enhanced significantly by judicious design of the inhomogeneity. The idea is to gain from an optimal combination of large pairing gap in the low doping regions and large carrier density in the highly doped ones\cite{Kivelson:2002}.

Enhancement of $T_c$ due to a similar mechanism was predicted in cuprate heterostructures composed of an underdoped  superconducting layer coupled to an overdoped metallic one.\cite{Berg:2008, Okamoto:2008,Goren:2009} The underdoped layer induces a proximity gap in the overdoped layer, which then contributes to the zero temperature phase stiffness of the system and considerably enhances it compared with the suppressed stiffness of the underdoped layer. On the other hand, the $d$-wave proximity gap which is induced on the metallic layer is small, and thus results in a sharp reduction of the stiffness with the temperature\cite{Goren:2009}. We found in Ref. \onlinecite{Goren:2009} that the combined effect can in principle lead to enhancement of $T_c$ compared with an optimally doped layer. However to attain such enhancement the coupling between layers needs to be much larger than the realistic coupling between the cooper-oxide planes. It is therefore unlikely that these simplified models provide a satisfactory explanation for the $T_c$ enhancement observed in various experiments on heterostructures.\cite{Yuli:2008,Gozar:2008,Jin:2011} However, if there is doping inhomogeneity within a plane the coupling between the overdoped and underdoped regions would naturally be large since they are connected by the in-plane rather than the c-axis tunneling. As we shall see this situation can indeed give rise to enhancement of the maximal critical temperature compared to a pure system.

Specific kinds of in-plane inhomogeneity and their effect on superconductivity have been previously investigated theoretically. For example a weak-coupling BCS theory of the
{\em attractive} Hubbard model showed that $T_c$ can be enhanced by periodic modulations of the weak attraction.\cite{Martin:2005} A density matrix renormalization group (DMRG) study of the {\em repulsive} Hubbard model on a two leg ladder showed that modulations of the hopping matrix element along the ladder can enhance the pairing correlations and thereby possibly increase the $T_c$ of a coupled ladder system.\cite{Karakonstantakis:2011} A direct study of the two dimensional Hubbard model using contractor renormalization (CORE) also indicated that there is an optimal modulation of the hopping matrix element, which maximizes the pairing correlations.\cite{Baruch:2010} Finally, dynamical mean field and cluster Monte Carlo calculations find increased pairing gap, and possibly $T_c$, in a state with charge modulation near $1/8$ doping.\cite{Maier:2010,Okamoto:2010}

The above studies focus on the effect of periodic commensurate charge modulations on the pairing order parameter. We complement and extend the analysis in several ways.
First, we use an effective theory, amenable to analytic treatment that allows to identify the physical origin of the various effects. Second we compute the temperature dependent superfluid stiffness, which at least in the underdoped cuprates is a more complete measure of superconductivity than the pairing amplitude and allows us to directly estimate $T_c$. Third, in addition to the stripe model treated in previous work we also consider random doping variations, which appear to be the
more generic situation in samples of doping above $1/8$. Both for the stripe model
and the random inhomogeneity we asses the possibility of enhancing $T_c$ by tuning the
magnitude of characteristic doping modulations and their length scale.

We implement the inhomogeneity in the form of inclusions of a highly overdoped phase, already in the metallic regime, embedded in a background of underdoped or optimally doped material. The case of mesoscopic inhomogeneity, where the metallic inclusions are of the size of the superconducting coherence length or larger is sketched in Fig. \ref{fig:overview}(a). This is treated within an effective stripe model of the metallic regions, where we average over stripe orientations to obtain an isotropic macroscopic stiffness. Another case we consider, is where the metallic regions are much smaller than the coherence length and are modeled as point impurities. This case is depicted in Fig. \ref{fig:overview}(b).

In both cases we include the crucial effects of strong coulomb repulsion and of the $d$-wave symmetry of the order parameter. The former is the reason for the low superfluid density $\rho_s$ at low doping, while the second  is responsible for the linear suppression of $\rho_s$ with $T$ at low temperatures.\cite{Lee:1997} These effects are taken into account within a slave boson mean field theory of the $t-J$ model.\cite{Zhang:1988,Kotliar:1988} Furthermore, we include Fermi-liquid-like corrections phenomenologically, to the description of low energy quasiparticles.\cite{Millis:1998,Wen:1998,Paramekanti:2002}

Regardless of the model for the metallic regions we find an increase of the zero temperature stiffness and for a wide range of doping levels, also higher critical temperature compared to the pure system with the same average doping. Furthermore, in the case of microscopic impurities we even predict that a higher $T_c$ can be attained even compared to the maximal $T_c$ at optimal doping of the pure system.
 
The paper is structured as follows: In Sec. \ref{section:overview} we give a general overview of the models used, of the assumptions that underlie our choice of models, and of the main results obtained in the different regimes.  Section \ref{section:stripes} gives a detailed treatment of a model representing mesoscopic inhomogeneity, while in section \ref{section:disorder} we consider a model with point impurities. Section \ref{section:summary} is a summary and discussion of the results.

\section{Overview}\label{section:overview}
%%%%%%%%%%%%%%
%%%%%% figure:overview
%: fig:overview
\begin{figure}[t]
\begin{center}
\includegraphics{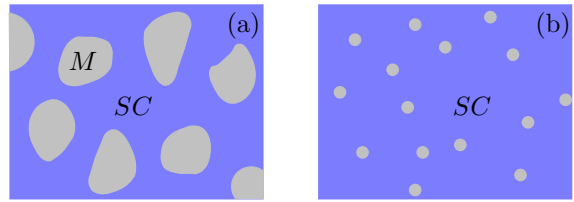}
\end{center}
\caption{(a) An illustration of a mesoscopic-scale inhomogeneous layer. The typical size of the metallic regions is equal or larger to the superconducting coherence length. (b) Microscopic-scale inhomogeneous layer. the metallic regions are point-like impurities. }
\label{fig:overview}
\end{figure}
%%%%%%%%%%%%%%%%%%%%%%%
In this section we introduce the framework for treating the inhomogeneous cuprate layer within a slave boson mean field theory. We describe the essential ingredients of the theory for the case of  mesoscopic inhomogeneity as well as for point impurities. Finally we summarize the main results that are derived in detail in later sections.

In order to describe doping inhomogeneity in cuprate materials we make use of models that can account for the effects of doping of the Mott insulating parent compound.
A simple theoretical framework that captures many of the important effects is the renormalized mean field theory (RMFT)\cite{Zhang:1988} or slave boson mean field theory (SBMFT)\cite{Kotliar:1988} of the $t-J$ Hamiltonian,
\be
H_{\rm tJ}=-P_G\sum_{ij\s}t_{ij}c_{i\s}^\dag c_{j\s} P_G+ h.c + \sum_{\langle ij\rangle}J_{ij}{\bf S}_i {\bf S}_j.
\ee
Here $J_{ij}=4t_{ij}^2/U$ is the super-exchange interaction, ${\bf S}_i=c\yd_{is}{\bf\s}_{ss'}c\nd_{is'}$ and $P_G=\Pi_i (1-n_{i\ua}n_{i\da})$ implements the Gutzwiller constraint, which prohibits double occupancy of sites.

The standard mean field treatment of the $t-J$ model includes two approximations. The first is to account for the projection only through renormalization of the hopping $t_{ij}\to g_{ij} t_{ij}$, while working in the full rather than the projected Hilbert space \cite{Zhang:1988}. The second approximation consists of a standard decoupling of the quartic term in both the Fock and BCS channels. The resulting mean field Hamiltonian is given by
\bea\label{HRMFT}
H_{\rm MF}&=&-\sum_{i,j,\s} (g_{ij}^t t + \chi_{ij} ) c_{i\s}^\dag c_{j\s}+h.c -\sum_{i,s} \mu_i c_{i\s}^\dag c_{i\s} \nn\\
&+& \sum_{\langle ij\rangle } \D_{ij}c_{i\ua}^\dag c_{j\da}^\dag + h.c
\eea
where $\chi_{ij}=3J_{ij}\sum_\s \langle c_{i\s}^\dag c_{j\s} \rangle/4$, $\D_{ij}=3J_{ij}\langle c_{i\ua}^\dag c_{j\da}^\dag-c_{i\da}^\dag c_{j\ua}^\dag\rangle/4$ and
$g_{ij}$ are doping dependent renormalization factors that account for the effect of the no-double-occupancy constraint. In a uniform system of doping ${p}$, $g=2{p}/(1+{p})$, $\chi_{ij}=\chi$ for all nearest neighboring $i,j$, and $\D_{i,i+\hat{x}}=-\D_{i,i+\hat{y}}=\D$ such that the pairing has a $d_{x^2-y^2}$ symmetry.

The mean field theory of the $t-J$ model captures the crucial fact that the zero temperature superfluid stiffness of underdoped cuprates scales linearly with the hole doping, $\rho_0\propto p$.\cite{Uemura:1989,Lee:1997} It also accounts for the $d$-wave symmetry of the gap that gives rise to a low energy quasiparticle spectrum of the form $E_{\bf k}=[({\bf v}_f {\bf k}_{||})^2+ ({\bf v}_\D{\bf k}_{\perp})^2 ]^{1/2}$. This form of the spectrum explains the observed linear reduction of the superfluid stiffness with temperature, $\rho_s(T)=\rho_0-b_0 T$ with $b_0=2\log{2} (2\sqrt{2}Zt)^2/(\pi v_f v_\D)$.\cite{Lee:1997}
However, the mean field theory does not give the correct value of $Z$. This can be viewed as a Fermi liquid correction that may be strongly renormalized at low energies due to quasi-particle interactions not included in the mean field theory.\cite{Millis:1998,Wen:1998,Paramekanti:2002} Therefore $Z$ is best taken as a phenomenological parameter to be extracted from experiments.\cite{Wen:1998,Ioffe:2002}

In this paper we extend the analysis of the stiffness and the critical temperature to
the case of an inhomogeneous system. Specifically we describe an underdoped system in the bulk ($0.1<p_1<0.15$) embedded with highly overdoped  metallic regions. We consider two regimes of inhomogeneity as illustrated in Fig.~\ref{fig:overview}. First is when the metallic regions are of the order or larger than the superconducting coherence length and second when they are of the order of one lattice constant. As discussed in the introduction a pertinent question we wish to address is whether such inhomogeneity can lead to enhanced $T_c$.

\subsection{Mesoscopic Inhomogeneity}
In the first model, described in Sec.~\ref{section:stripes}, we assume a 2D mixture of a superconducting underdoped phase (of doping $p_1$) and an extremely overdoped metallic phase (of doping $p_2>0.3$).
The doping level varies considerably only across a length scale of the order of the coherence length $\xi\sim v_f/\D$, which is typically around 5 lattice spacings, such that the 2D regions are of intermediate size $\ge \xi$ as depicted in Fig.~\ref{fig:overview}(a). This scenario is reminiscent of various experiments that find gap variations on a similar scale, of $5-10 nm$ \cite{Chang:1992,Pan:2001,Howald:2001,Gomes:2007,Parker:2010}.

We model this system as a mixture of striped domains, each one with alternating underdoped and overdoped stripes along the $x$ or $y$ direction, such that on a macroscopic scale the system is fourfold rotationally invariant [see Fig.~\ref{fig:inhomsketch}(a)]. This allows us to obtain an expression for the superfluid stiffness of the entire system. The superconducting stripes are described by the $t-J$ Hamiltonian and the metallic stripes are modeled by free fermions. We vary the widths of the stripes in order to explore the superconducting properties in various geometries. To calculate the critical temperature of the inhomogeneous mixture, we solve self consistently the Bogoliubov de Gennes equations for Hamiltonian (\ref{HRMFT}) allowing for position dependent $g_{ij}$, $\D_{ij}$ and $\chi_{ij}$.
We derive a general formula for the superfluid stiffness $\rho_s(T)$ of a striped superconductor in terms of response kernels that can be directly calculated from the Bogoliubov de Gennes solution [see Eqns. (\ref{Kyy}),(\ref{Kxx})].
We then use the Kosterlitz-Thouless criterion $\rho_s(T_c)=2T_c/\pi$ to determine $T_c$ of the mixed system.

We show that there exist optimal configurations which allow for an enhanced zero temperature superfluid stiffness in the inhomogeneously doped layer, compared with the homogeneous superconducting one. This is a consequence of proximity effect that leads to a gap in the metallic regions. The metallic regions, having a large density of charge carriers, can then contribute significantly to the superfluid stiffness of the inhomogeneous layer at $T=0$. On the other hand, since the proximity gap is much smaller than the original superconducting gap, the reduction of the stiffness at finite temperature is sharper than in the uniform superconductor.
It therefore does not immediately follow that the interplay of these two effects can lead to an enhancement of the critical temperature. Previously  we found that such an enhancement is possible in a bilayer of underdoped and overdoped cuprates, under appropriate conditions\cite{Goren:2009}.
In the present scenario, however, we find that $T_c$ of the inhomogeneously doped layer
is lower than the one of a homogeneous underdoped superconductor of doping $p_1$. The reason is that already at $T=0$ the enhancement of the stiffness due to enlarged carrier density is counteracted to a large extent by a significant paramagnetic suppression of the stiffness which is inevitable in inhomogeneous superconductors. Consequently, the zero temperature stiffness is enhanced compared with the uniform case, but not enough to allow for an enhancement of $T_c$.

Nonetheless, we find that the critical temperature of the system increases with the reduction of the relative width of the metallic stripes. This allows for a large proximity gap in the metallic regions, manifested in a relatively small reduction of the stiffness at finite temperature. In order to maximize the proximity effect, but  at the same time allow for a significant contribution of carriers from the metallic region, an optimal configuration should have small but relatively dense metallic regions. In the following we consider the effect of small metallic regions.

\subsection{Microscopic Inhomogeneity}
In this model, described in Sec.~\ref{section:disorder} we assume microscopic overdoped regions (doping $p_2$)  which are placed in a low doping superconducting background (doping $p_1$), see Fig.\ref{fig:overview}(b).
The microscopic overdoped regions are modeled as single site impurities with zero or very weak local Hubbard repulsion ($U\sim0$), which induces modified hopping and pairing amplitudes along their neighboring bonds, as depicted in Fig.~\ref{fig:dis-sketch}. The hopping amplitude along these bonds is the bare $t$ rather than the renormalized value of SBMFT, and the local pairing strength there is suppressed to zero.

In the presence of the bond disorder we compute the temperature dependent superfluid stiffness using a perturbative expansion to second order in the impurity strength for disorder averaging (second order Born approximation).
Then we determine the transition temperature using the Kosterlitz-Thouless criterion as before.

 Since the variations in doping generates unconventional bond disorder, the calculation bears several important differences from the standard impurity averaging. The most important difference is that the bond disorder introduces local modulations in the current operator thus renormalizing the coupling to the external vector potential. As a result, the superfluid response obtains vertex corrections which have no counterpart in standard (on-site) impurity averaging but play a crucial role in our case. One important effect of these corrections is to allow for an enhancement of the zero temperature diamagnetic stiffness of the disordered system compared with the pure one.
 A second effect of the vertex corrections is to introduce a paramagnetic reduction of the stiffness at zero temperature, similarly to the mesoscopic inhomogeneous scenario.
 In addition, the disorder introduces self-energy corrections which amount to an anti-proximity effect that acts to reduce the average pairing gap and contributes to the suppression of the stiffness at finite temperature.

 The net effect that we find is an enhancement of the superfluid stiffness and a concomitant increase in the critical temperature for a given bulk doping level $p_1$. Interestingly, we even find an overall enhancement of the maximal $T_c$, that is at optimal doping, compared to the maximal $T_c$ of the homogeneous system.

\section{Mesoscopic scale inhomogeneity}\label{section:stripes}
\subsection{The Model}\label{subsec:Model1}
\begin{figure*}[t]
\begin{center}
\includegraphics{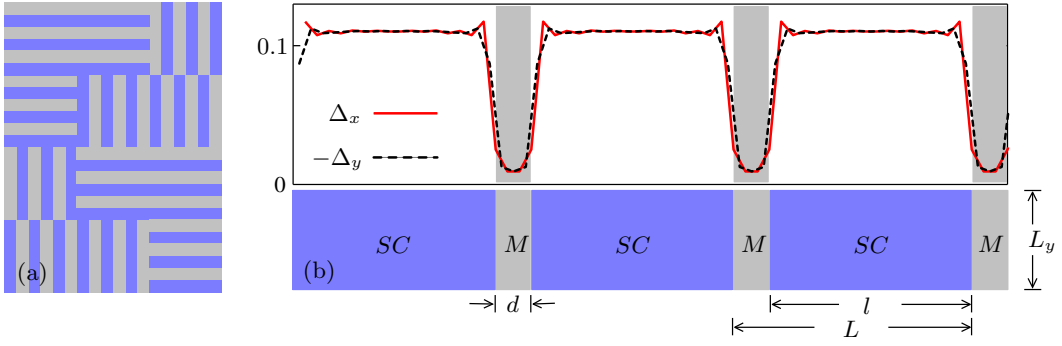}
\end{center}
\caption{(a) A model of the inhomogeneous layer as an array of striped domains, which on average is macroscopically fourfold rotationally invariant. (b) The self consistent gap profile, solved for $J=t/3, t'=0, x=0.1, d=3a$. $SC$ and $M$ denote the underdoped and overdoped regions respectively.}
\label{fig:inhomsketch}
\end{figure*}

%%%%%%%%%%%%%%%%%%%%%%%%%%%%%%%%%%%%%%%%
In this section we consider a stripe model. The inhomogeneity is of mesoscopic scale in the sense that the width of the stripes is of the order or somewhat larger than the coherence length associated with the superconducting regions.
The superfluid response of such a striped system is of course anisotropic. However we envision that  it becomes isotropic on macroscopic scales due to mixing of striped domains with random orientations as sketched in Fig~\ref{fig:inhomsketch}(a). The doping level of the stripes alternates between $p_1 $ in underdoped superconducting (SC) stripes of width $l$, and $p_2$ in metallic (M) stripes of width $d$.

As the Hamiltonian of a single domain we take the $t-J$ model
\bea\label{H:stripes}
H&=&P_G\sum_{i,j,\s}  t_{ij}c_{i\s}^\dag c_{j\s} P_G + \sum_{\langle ij\rangle } J_{ij}{\bf s}_i {\bf s}_j
\eea
where $P_G$ is the Gutzwiller projection that eliminates double occupancy of sites in the superconducting stripes, but does not affect the metallic stripes. The magnetic exchange coupling is $J_{ij}=J$ in the superconducting stripes and it vanishes in the metallic stripes.

%%%%%% figure:inhomsketch
%: fig:inhomsketch
%%%%%%%%%%%%%%%%%%%%%%%
We treat the space dependent projection and exchange interaction within slave boson mean field theory (SBMFT) \cite{ Kotliar:1988,Zhang:1988}. The resulting Hamiltonian is of the form (\ref{HRMFT}), with space dependent $\mu_i,g^t_{ij}$, $\chi_{ij}$ and $\D_{ij}$.
The electro-chemical potential $\mu_i$ is determined such that the doping levels of the superconducting and the metallic regions are $p_1$ and $p_2$ respectively. Due to the spatial variations in doping the renormalization of the hopping varies in space too and equals $g^t_{ij}={2p_1}/(p_1+1) $ in the superconducting stripes and $g^t_{ij}=1$ in the metallic stripes, while the tunneling at the interface between the two regions is renormalized by $g^t_{ij}=\sqrt{{2p_1}/(p_1+1)}$).

Given all the parameters of the mean field model, the fields $\chi_{ij}$ and $\D_{ij}$ can now be determined by the self consistency conditions:
\bea\label{eqn:selfconsistent}
 \chi_{ij}&=&\frac{3J_{ij}}{8}\sum_\s \langle c_{i\s}^\dag c_{j\s}  \rangle \nn\\
\D_{ij}&=&\frac{3J_{ij}}{8}\langle c_{i\ua}^\dag c_{j\da}^\dag - c_{i\da}^\dag c_{j\ua}^\dag \rangle
\eea
 An example of the resulting profile of the pairing amplitudes is plotted in Fig. \ref{fig:inhomsketch}(b), where $\D_x$ and $\D_y$ denote the pairing amplitudes on bonds along the $x$ and the $y$ directions respectively.
Because the pairing amplitude in the metallic regions is non zero, these regions contribute to the superfluid stiffness at low temperatures.

\subsection{Calculation of the Superfluid Stiffness}\label{subsec:calculation1}
%%%%%%%%%%%%%%%%%%%%%%%%%%%%%%%%%%%%%%%%
In a striped system the superfluid response depends on the direction of the applied phase twist. However, we assume that the system consists of many striped domains with random orientations along the principal axes. Under this assumption the superfluid response is homogeneous on large scales. It was shown in Ref.~\onlinecite{Carlson:2000} that the superfluid stiffness of the mixed domains is given by the geometric mean of the $x$ and $y$ components of the stiffness of a single domain $\rho_s=\sqrt{K_{xx}K_{yy}}$. Here $K_{aa}$ ($a=x,y$) are the diagonal components of the response tensor,
 \be\label{stiffness:definition}
 K_{ab}=\frac{I_a}{\D \theta_b}
 \ee
 where $\D \theta_b$ is the static phase difference applied across the system in the $\hat{b}$ direction and $I_a=\int  J_a ds_a$ is the total current measured in the $\hat{a}$ direction.

 In an inhomogeneous system, we express the stiffness tensor using the microscopic response kernel $\kappa_{ab}(r,r')$ defined through
 \be\label{localresponse:definition}
J_a(r)=\int_{t<t'}dr' \ \kappa_{ab}(r,r')\partial_b \theta(r').
 \ee
The response kernel can then be calculated using the standard Kubo formalism.
In the $y$ direction, parallel to the stripes, the stiffness $K_{yy}=\int dx J_y/\D \theta_y$ is simply an algebraic sum of the response kernels along the $x$ direction
\be\label{Kyy}
K_{yy} =  \int dx \int dx' \kappa_{yy}(x,x',q_y=0),
\ee
where we used the uniformity along the $y$ direction to express it in terms of the $q_y=0$ Fourier component of the response kernel.

To derive an analogous relation between $K_{xx}$  and $\kappa_{xx}(x,x')$, we follow the arguments presented in Ref.~\onlinecite{Carlson:2000}. It is convenient to use the lattice formulation and express all convolution integrals as matrix products.
The response kernel is then defined by (summation over repeating indices implied),
\be\label{localresponse:definition:lattice}
J_a(i)=\ \kappa_{ab}(ij)\partial_b\theta(j)
\ee
where the $y$ dependence is suppressed and we denote by $i,j$ the position in the $x$ direction only.

A static current is divergenceless $\nabla\cdot {\bf J}=0$ and therefore derived from a potential, $J_a(i)=\e_{ab}\partial_b \phi(i)$. Plugging this back into (\ref{localresponse:definition:lattice}) we obtain
\be\label{Phi:definition}
\e_{ab}\partial_b\phi(i)= \ \kappa_{ab}(ij)\partial_b\theta(j).
\ee
We can now derive a second relation between $\phi$ and $\theta$ if we multiply by $\e \kappa^{-1}$ from the left. Defining $\kappa^D_{ab} = \e_{ac} \kappa^{-1}_{cd}\e_{db}$ we arrive at
\be\label{derivedrelation}
\kappa^D_{ab}(ji)\partial_b\phi(i)=\e_{ab}\partial_b\theta(j).
\ee
Equations (\ref{Phi:definition}) and (\ref{derivedrelation}) establish a duality relation $\theta \leftrightarrow \phi$ and $\kappa \leftrightarrow \kappa^D$. We make use of this duality in the calculation of $K_{xx}$.
The response $K_{xx}$ is obtained by applying a phase difference $\D\theta_x$ along the $x$ direction and measure the resulting current $J_x$, equivalent to a difference in $\phi$ in the transverse direction, $\D \phi_y=\int dy J_x$. The response is then $K_{xx}=\D \phi_y/\D\theta_x$.

Using Eqn.~(\ref{derivedrelation}) we deduce that $\partial_x\theta=-\kappa_{yy}^D\partial_y \phi$, which allows us to apply relation (\ref{Kyy}) with $\theta \leftrightarrow \phi$ and $\kappa\leftrightarrow \kappa^D$. Doing so, we obtain the response $\D \theta_x/\D\phi_y=-\int dx dx' \kappa^D_{yy}=K_{xx}^{-1}$. This gives the result
\be\label{Kxx}
K_{xx}[\kappa] = \frac{1}{\int dx\int dx'\kappa^{-1}_{xx}(x,x')}.
\ee
When the stripes are macroscopic we can take the response functions to be translationally invariant within a stripe. Then (\ref{Kxx}) reduces to the well known fact that  the stiffness of macroscopic objects in series adds like resistors in parallel.

The Superfluid stiffness is now expressed in terms of the local response Kernel which can be computed using the standard Kubo formalism. The diamagnetic contribution of the response to a transverse vector potential is
\be\label{kernel:dia}
 \kappa_{aa}^{\rm dia} (x,x',q_y \ra 0)=e^2\delta_{xx'}\langle -K_a(x,q_y \ra 0)\rangle.
 \ee
 where $K_a(j)=-t(j)\sum_\s( c_{j+\hat{x}_a,\s}^\dag c_{j,\s} + c_{j,\s}^\dag c_{j+\hat{x}_a,\s})$
  is the local kinetic energy operator. The paramagnetic contribution is
 \bea\label{kernel:para}
 \mathcal{\kappa}_{ab}^{\rm para}(x,x',q_y,i\omega_n)&=&-\int_0^\b d\tau e^{i\omega_n \tau}\times\\
 &\times& \langle T_\tau j_a^p(x,q_y ,\tau)  j_b^p(x',-q_y,0)\rangle \nn,
 \eea
 in the limit  $\omega=0, q_y\ra0$. Here $ j_a^p$ is the paramagnetic current operator:
\be
  j_a^p(j)=it(j) \sum_\s\left( c_{j+\hat{x}_a,\s}^\dag c_{j,\s} - c_{j,\s}^\dag c_{j+\hat{x}_a,\s} \right).
 \ee
 In order to calculate (\ref{kernel:dia}) and (\ref{kernel:para}) we diagonalize the Hamiltonian using the Bogoliubov transformation
 \bea\label{bogoliubov}
c_{j\ua}^\dag(k_y) &=& \sum_\a u^*_{\a}(j,k_y)\g_\a^\dag(k_y) \nn\\
  c_{j\da}(-k_y) &=& \sum_\a v^*_{\a}(j,k_y)\g_\a^\dag(k_y).
 \eea
 We then solve the self consistent equations (\ref{eqn:selfconsistent}) and express $K_a(j)$ and $ j_a^p$ in the new basis.
 As an example, the $x$ component of the static paramagnetic response kernel is
\bea\label{xx-kernel:result}
\kappa_{xx}^{\rm para}(j,j', q_y)&=&\sum_{k_y\a\b} \mathcal{J}_{\a\b}^{(x)}(j,k_y,q_y) \mathcal{J}_{\b\a}^{(x)}(j',k_y,-q_y)\times\nn\\&\times&\mathcal{P} \left[ \frac{f(E_{\a,k_y})-f(E_{\b,k_y-q_y})}{E_{\a,k_y}-E_{\b,k_y-q_y}}\right]
\eea
where $\mathcal{P}$ denotes the principal part, $E_{\a,k_y}$  are the eigen-energies, $f(\e)=[1+e^{\b\e}]^{-1}$ is the Fermi-Dirac distribution, and
\bea
\mathcal{J}^{(x)}_{\a\b}(j,k,q)&\equiv& i t(j)
\bigl[
u^*_{\a}(j+1,k)u_{\b}(j,k-q) \nn\\
&+&v^*_{\a}(j+1,k)v_{\b}(j,k-q)
\bigr]
 + \rm{c.c.}\nn
\eea
To solve the periodic problem we introduce an additional superlattice momentum, whose index is suppressed here for simplicity.

Finally, the superfluid stiffness $\rho_s(T)=\sqrt{K_{xx}K_{yy}}$ is computed from the response kernels $\kappa_{aa}(x,x')= \kappa_{aa}^{\rm dia}(x,x') + \kappa_{aa}^{\rm para}(x,x')$ using Eqns. (\ref{Kyy}) and (\ref{Kxx}).

\subsection{Results and Discussion}\label{subsec:results1}
%%%%%%%%%%%%%%%%%%%%%%%

%%%%%% figure:rho_Tc
%: fig:rho_Tc
\begin{figure*}[t]
\begin{center}
\includegraphics{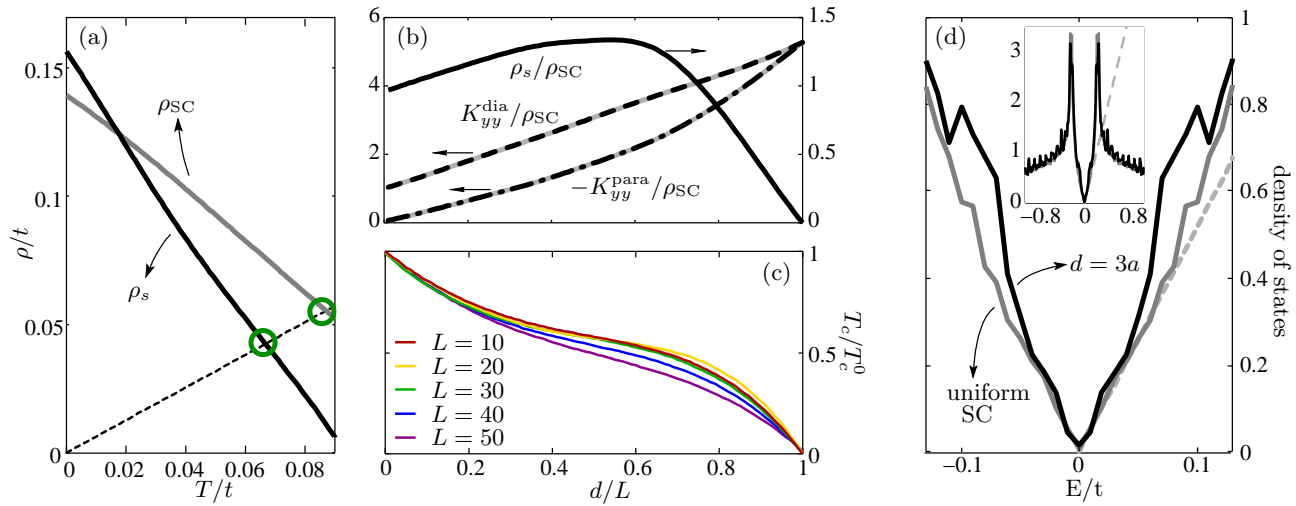}
\end{center}
\caption[stiffness and Tc - stripes]{ (a) Temperature dependent superfluid stiffness of a striped versus a homogeneous superconducting layer. In this example the width of metallic stripes (doping $p_2=0.35$) in this example is $d=3$ and that of the superconducting stripes ($J=t/3$ and doping $p_1=0.1$) is $l=17$ (in lattice constants). The critical temperature estimated from the Kosterlitz-Thouless criterion, is seen to be higher in the homogeneous system (circle marks).
(b) Zero temperature stiffness vs the relative size of the metallic segments, $d/L$ for  a fixed $L=20$. The diamagnetic (dashed) and paramagnetic (dot-dashed) contributions to the stiffness $K_{yy}$ are plotted for comparison. All results are normalized by the zero temperature stiffness $\rho_{\rm SC}$ of the uniform superconductor. (c) Critical temperature, normalized by the uniform value, vs $d/L$ for the same parameters as in (b) and different unit cells $L$ (in units of the lattice spacing $a$). (d) The density of states for a system with the same parameters as (a). At low energies the density of states (DOS) of the striped system (black) is identical to the DOS of a uniform underdoped superconductor of the same size (gray). The dashed line denotes the calculated slope $d\nu/dE$ at $E=0$ for the uniform superconductor. }
\label{fig:rho_Tc}
\end{figure*}
%%%%%%%%%%%%%%%%%%%%%%%

We first discuss the superfluid stiffness at zero temperature, and then turn to an analysis of the temperature dependence of the stiffness, in order to estimate the critical temperature of the striped system.

 The superfluid stiffness at zero temperature $\rho_s(T=0)$ is plotted in Fig.~\ref{fig:rho_Tc}(b)
as function of the relative width of the metallic stripes, $\zeta=d/(d+l)$ at fixed width $l+d=20a$ ($a $ is the lattice spacing). The doping levels of the superconducting and the metallic stripes are $p_1=0.1$ and $p_2=0.35$ respectively.
Note the enhancement of the stiffness compared with the uniform superconductor $\rho_{\rm SC}$ which is maximized for $\zeta\sim 0.5$.

The enhancement of the zero temperature stiffness and the optimal volume fraction $\zeta$ are determined by the interplay of two effects. First, the metallic regions are gapped by the proximity effect, and contribute their large number of carriers  to the diamagnetic superfluid density $K_{yy}^{\rm dia}$ which increases with $\zeta$. However, this increase is partially countered by a zero temperature paramagnetic term $K_{yy}^{\rm para}$ special to inhomogeneous superconductors. A similar effect was noted by us in a bilayer heterostructure \cite{Goren:2009}.

To see if the moderate net increase of the zero temperature stiffness will facilitate enhancement of the transition temperature we compute the full temperature dependence of the stiffness. As an example Fig.~\ref{fig:rho_Tc}(a) shows the result for a specific ratio $d/(d+l)=3/20$ with $p_1=0.1$ and $p_2=0.35$.
 The macroscopic stiffness $\rho_s$ is seen to decrease approximately linearly with temperature, as in a uniform $d$-wave superconductor, but with a larger slope $d\rho_s/dT$. As a result, the transition temperature, determined using the Kosterlitz-Thouless criterion $\rho_s(T_c)=2T_c/\pi$, is found to be lower in the inhomogeneous layer despite the increased stiffness at zero temperature. This remains the case in all possible stripe geometries, as shown in Fig.~\ref{fig:rho_Tc}(c).

The slope $d\rho_s/dT$ is affected by two main factors: the first is the density of states (DOS) of low energy quasiparticles that carry the paramagnetic current and the second is the effective charge of these quasiparticle (or the effective current renormalization). The DOS of the system is plotted in Fig.~\ref{fig:rho_Tc}(d). Below a threshold energy of $E\sim 0.05 t$ the DOS is the same as in the uniform superconductor of $p_1=0.1$. This agrees with experimental results in inhomogeneous cuprate superconductors \cite{Howald:2001}. Note that the limit of very narrow metallic stripes ($d=3a$) preserves the low energy DOS up to a relatively high energy, compared with the critical temperature. As the metallic stripes get wider, there are more low energy states that contribute to the reduction of the stiffness with the temperature.

 Despite the fact that the low energy DOS is the same as in the uniform superconductor, the slope $|d\rho_s/dT|$ is still larger than in the uniform case. This is a consequence of the difference in the effective charge of quasiparticles in the two systems: In the underdoped superconducting regions the quasiparticle charge is renormalized down by a factor of $g_t=2p_1/(1+p_1)$, whereas in the metallic regions there is no such renormalization and the current is carried by electrons. As a result, at finite temperature the stiffness reduction in the inhomogeneous system is steeper than in the uniform underdoped superconductor.

 Here we should note again that, in general, the renormalization of the current carried by a quasi-particle that enters the low temperature dependence of the stiffness is a Fermi-liquid parameter that may be  renormalized compared to the SBMFT value of $g_t=2p_1/(1+p_1)$. Indeed measurements of the temperature dependent stiffness give a renormalization factor is independent of doping over a wide range of doping in contradiction to the mean field prediction. However for an inhomogeneous system there is no unambiguous way to replace the mean field value of the current renormalization by a single phenomenological parameter. Moreover the existence of a length scale $d$ of the superconducting regions may introduce a cutoff that prevents this parameter from flowing far from its mean field value at low energies.

The stripes model shows that doping inhomogeneity on a mesoscopic scale can lead to an increased superfluid stiffness at zero temperature. This is a consequence of a proximity gap that opens in the metallic stripes which then contribute their high carrier density to the stiffness. On the other hand the metallic stripes also give rise to low energy states that  hasten the reduction of stiffness with temperature. In addition there is a paramagnetic reduction of the stiffness even at zero temperature due to the impurities. For these reasons the transition temperature of the striped system is found to be always lower than that of the homogeneous system.
The highest $T_c$ is obtained for the narrowest metallic stripes because then the proximity coupling to the bulk is high and the Andreev bound state are only slightly below the gap.
It is therefore tempting to consider the case of even smaller metallic regions by reducing the length of the stripes in addition to their width to a microscopic scale. In the following section we examine a model that takes a step in this direction.

\section{Microscopic scale inhomogeneity}\label{section:disorder}

\subsection{The Model}\label{subsec:Model2}
\begin{figure}[h]
\begin{center}
\includegraphics{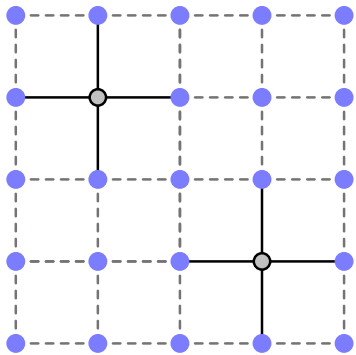}
\end{center}
\caption{\emph{Doping inhomogeneity on a microscopic scale.} An illustration of the model. Nearest neighbouring bonds to the impurity sites (solid lines) are characterized by enhanced hopping amplitude $gt+\delta_t$ and reduced pairing $\D_0+\delta_\D$, with respect to the superconducting background of $gt$ and $\D_0$ respectively.}
\label{fig:dis-sketch}
\end{figure}

In this section we consider a scenario in which the metallic regions embedded in the underdoped superconductor are nearly point like. We model these metallic impurities as cross vertices of the square lattice (see Fig. ~\ref{fig:dis-sketch}) on which the average doping $p_2$  is higher than the bulk average $p_1$. The effective hopping matrix elements and the pairing amplitudes on these links naturally take different values than the bulk. Specifically, in the mean field model of Eqn.~(\ref{HRMFT}) the parameters $g^t_{ij}$, $\D_{ij}$ and $\chi_{ij}$ take a different value on the impurity bonds.

We analyze two scenarios: in the first, the impurities are metallic, with doping $p_2>0.3$, such that on the impurity bonds $g_{ij}=1$ and $\D_{ij}=\chi_{ij}=0$. In the other scenario the excess doping on the impurity sites $p_2-p_1$ is small, leading to $g_{ij}=g(p_2)$ and $\D_{ij}=\D(p_2)$ with the doping dependence of SBMFT. In this case we assume that $\chi$, which has a very weak doping dependence, is uniform throughout the system.

%%%%%% figure:sketch
%: fig:dis-sketch
%%%%%%%%%%%%%%%%%%%%%%%

The Hamiltonian $H=H_0+H_{\rm imp}$ consists of a uniform part and an impurity contribution. Written in momentum space, the uniform Hamiltonian is the Fourier transform of (\ref{HRMFT}),
\be
H_0=\sum_{\bf k} \Psi_{\bf k}^\dag (\xi_{\bf k} \s_3 + \D_{\bf k} \s_1)\Psi_{\bf k}.
\ee
Here $\Psi_{\bf k}^\dag= \{c_{{\bf k }\ua}^\dag, \  c_{{\bf -k }\da}\}$, $\s_a$ are Pauli matrices, $\xi_{\bf k}=-2t_{\rm eff} (\cos{k_x}+\cos{k_y})-\mu$ and $\D_{\bf k}=\D_0 (\cos{k_x}-\cos{k_y})$ with $t_{\rm eff}=g(p_1)t+\chi$. The impurity Hamiltonian is
\be\label{disorderpotential}
H_{\rm imp}=\sum_{\bf kk'} \Psi_{\bf k}^\dag  \rho_{\bf k'-k} \left[{U}_{\bf kk'}\s_3  + {V}_{\bf kk'}\s_1 \right] \Psi_{\bf k'}
\ee
where
\bea
U_{\bf kk'}&=&-2\delta_t (\cos{k_x}+\cos{k_y}+\cos{k'_x}+\cos{k'_y})\nn\\
V_{\bf kk'}&=&\delta_\D(\cos{k_x}-\cos{k_y}+\cos{k'_x}-\cos{k'_y}) \nn\\
 \hat{\rho}_{{\bf k'}-{\bf k}}  &=& \sum_\a e^{-i({\bf k'}-{\bf k}){\bf r}_\a}
\eea
Here $\d_\D=\D(p_2)-\D(p_1)$, and $\{{\bf r}_\a\}$ are the impurity sites. The excess hopping at the impurity sites is $\d_t=t_{\rm eff}(p_2)-t_{\rm eff}(p_1)$. In the case of metallic impurities we set $t_{\rm eff}(p_2)=t$. 

The above terms result from the shift in doping level from $p_1$ to $p_2$ near the impurity. We should in principle include also the direct impurity potential, which caused the change in hole concentration. Such a potential that acts locally on the impurity as $U({\bf r})=U_0\d({\bf r}-{\bf r}_{\bf\a})$, can be regarded as a $k$ independent contribution to $U_{\bf kk'}$. 
The magnitude of this term can be estimated from the observed change in hole concentration through $U_0 \simeq (p_2-p_1)/\kappa$, where $\kappa$ is the local compressibility. 
We omit this term from the calculations described below. Then, at the end of  Sec.~\ref{subsec:results2} we quantify the contribution of the direct impurity potential and explain why it can be neglected.

Our goal is to compute the temperature dependent superfluid stiffness and estimate the transition temperature of the inhomogeneous layer compared to a uniform layer. To this end we use the Born approximation to perform the disorder average.
This is strictly valid in the limit of dilute uncorrelated impurities and weak disorder. We expand to first order in the impurity concentration $n_i$ and second order in the strength of a single impurity $\delta_t/t_{\rm eff}$ and $\delta_\D/\D_0 $. In practice we will allow $\d_\Delta/\D_0$ to be close to $-1$ which is the case when the overdoped inclusions are already in or close to the metallic regime.

\subsection{Calculation of the Superfluid Stiffness}\label{subsec:calculation2}

The stiffness is the linear response of the system to an externally applied vector potential ${\bf A}({\bf r})$.
In order to calculate it in the disordered system, it is convenient to resort to the real space Hamiltonian (\ref{HRMFT}) and include a vector potential through a Peierls substitution, $g_{ij}t\ra g_{ij}t\exp{[ieA_{ij}]}=g_{ij}t\exp{[ieA_x({\bf r}_i)]}$ in the case of a vector potential in the $x$ direction.
For the linear response calculation we expand the Hamiltonian to second order in $A_x$ \cite{Scalapino:1993},

\be
H(A_x)=H(0)-\sum_{\bf r} \left[e j_x({\bf r})A_x({\bf r}) + \frac{e^2}{2}K_x({\bf r}) A_x^2({\bf r})\right]
\ee
with
\bea\label{jK}
j_x({\bf r})&=&i\sum_{{\bf r},\s} t_x({\bf r})(c_{{\bf r}+x,\s}^\dag c_{{\bf r},\s}-c_{{\bf r},\s}^\dag c_{{\bf r}+x,\s})\nn\\
K_x({\bf r})&=&-\sum_{{\bf r},\s} t_x({\bf r})(c_{{\bf r}+x,\s}^\dag c_{{\bf r},\s}+c_{{\bf r},\s}^\dag c_{{\bf r}+x,\s})\nn.
\eea
Here $ t_x({\bf r})=g(p_1)t+\delta_t^{c}\sum_\a \d({\bf r}-{\bf r}_\a)$ is the coupling to the external vector potential, in the presence of the modified bonds around sites ${\bf r}_\a$. The excess local current on the impurity sites is $\delta_t^{c}=t[g(p_2)-g(p_1)]$. In the case of highly overdoped impurities ($p_2>0.3$) we take $g(p_2)=1$. Note that this impurity contribution is different from $\d_t$ that appears in the impurity Hamiltonian (\ref{disorderpotential}). The reason is that the external vector potential couples only to the hopping $g_{ij}t$, and not to the Fock term proportional to $\chi$, which originated from the magnetic exchange interaction.

\begin{figure}[t]
\begin{center}
\includegraphics{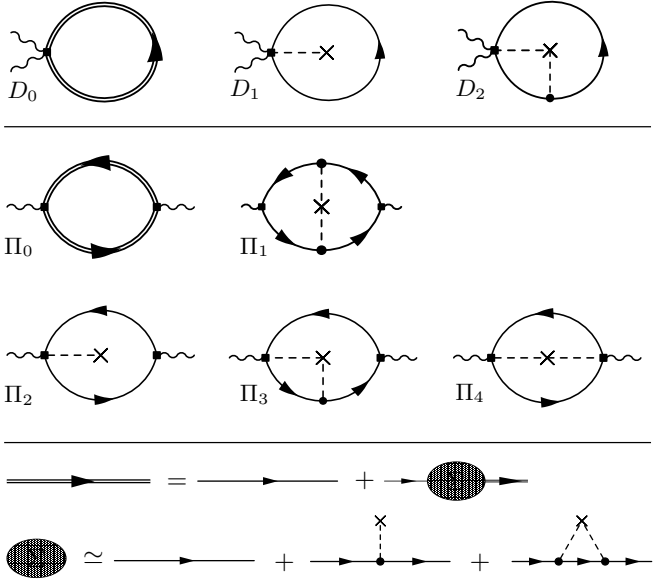}
\end{center}
\caption{\emph{Diagrammatic calculation of the superfluid stiffness in the disordered system.} Top: diamagnetic contributions. Middle: Paramagnetic contributions. Bottom: Green's function renormalization and the definition of the self energy within Born approximation. A dashed line corresponds to a scattering event and $\times$ denotes a single impurity. Note that the scattering is a matrix in Nambu space.}
\label{fig:diagrams}
\end{figure}

The superfluid stiffness is now given by\cite{Scalapino:1993}
\be
\rho_s=\overline{\langle -K_x \rangle} +\lim_{{\bf q}\ra 0} \overline{\Pi_{xx}}({\bf q},\omega=0)
\ee
where $\overline{X}$ denotes the average over disorder realizations and,
\bea
 \overline{\Pi_{xx}}({\bf q},\omega_n)\! &\!=\!&\!-\!\int_0^\b d\tau e^{i\omega_n\tau} \overline{\langle j_x({\bf q},\tau) j_x({\bf -q'},0)\rangle} \nn
\eea
Note that after disorder averaging the RHS is proportional to $\delta_{\bf qq'}$.
The different contributions to $\rho_s$ are presented as diagrams in Fig.~\ref{fig:diagrams}, where we denote diamagnetic terms by $D_\alpha$ and paramagnetic terms by $\Pi_\a$.

One type of correction to the stiffness stems from standard renormalization of the electron self-energy by the impurities. Such corrections are given by diagrams $D_0$ and $\Pi_0$ in Fig.~\ref{fig:diagrams}. Similar terms would arise in the common case of point (on-site) impurities. We note that the vertex correction $\Pi_1$ vanishes due to inversion symmetry of the impurity potential.

A second type of correction to the stiffness is special to the bond disorder we consider here. The disorder in the hopping amplitude introduces modulations in the local current operator and kinetic energy, proportional to $\d_t$. This causes a direct renormalization of the coupling to the external vector potential, as represented in diagrams $D_1, D_2, \Pi_2,\Pi_3$ and $\Pi_4$ in Fig.~\ref{fig:diagrams}.\\

%%%%%%%%%%%%%%%%%%%%%%%%%%%%%%%%%%%%%%%%%%%%%%
%%%%%%%%%%%%%%%%%%%%%%%
%%%%%% figure:diagrams
%: fig:diagrams
%%%%%%%%%%%%%%%%%%%%%%%

 {\bf Self energy corrections.} The disorder in the hopping and pairing strength introduces renormalizations to the spectrum parameters or to the electronic Green's function, which in turn, affect the superfluid stiffness. Such corrections are represented by diagrams $D_0$ and $\Pi_0$ in Fig.~\ref{fig:diagrams}. In order to calculate these diagrams we first compute the renormalized Green's function using the Born approximation.

The Dyson equation for the disorder averaged Green's function is depicted in Fig.~\ref{fig:diagrams} and given by
\be
G_{{\bf k},\omega_n}^{-1}=(G^0_{{\bf k},\omega_n})^{-1} + \Sigma_{{\bf k},\omega_n},
\ee
where the bare Green's function is
\be
(G^0_{{\bf k},\omega_n})^{-1}  = i\omega_n\s_0 - \xi_{\bf k} \s_3 - \D_{\bf k}\s_1,
\ee
and $\Sigma_{{\bf k},\omega_n}\equiv \sum_{a=0}^3\Sigma_a\s_a $ is the self-energy after disorder averaging.
To calculate the self energy explicitly in the limit of small $\omega_n$, we use the fact that the main contributions arise from the vicinity of the nodal points $\D_{\bf k}=\xi_{\bf k}=0$. We expand around these points and solve self consistently for the decay rate $\Sigma_0$ in the limit of $\omega\ra0, {\bf k}\ra {\bf k}_{\rm node}$ similarly to Ref.~\onlinecite{Lee:1993}. The high energy cutoff for this approximation is defined as $p_0$. This calculation gives (see Appendix \ref{app:selfenergy}) ,
\bea\label{selfenergy}
\Sigma_0&\simeq& -ip_0 e^{-\frac{2 \pi v_f v_\D t_{\rm eff}^2}{n_i(2\mu\delta_t)^2}}
\\
\Sigma_1({\bf k})&\simeq& 2n_i \frac{\delta_\D}{\D_0} \left(1-\eta\right)\D_{\bf k}\nn\\
\Sigma_3({\bf k})&\simeq&\nn  2n_i \frac{\delta_t}{t_{\rm eff}}\left(1-\eta\right)\xi_{\bf k} + \delta_\mu \nn
\eea
where $\eta\equiv S_1 {\delta_\D}/{\D_0} +(1-S_1){\delta_t}/{t_{\rm eff}} $, $S_1\equiv \sqrt{2\D_0/t_{\rm eff}}/\pi=2\sqrt{v_\D/ v_f}/\pi$, and $\delta_\mu$ is a ${\bf k}$ independent constant that renormalizes the chemical potential.

The low energy limit of $\Sigma_0$ is exponentially small close to zero doping, and is further suppressed by the large number $t_{\rm eff}^2/(n_i \delta_t^2)$. We solve  for the other components of the self energy under the self consistent assumption that any ${\bf k}$ dependence  of $\Sigma_0$ is negligible and indeed get that the entire effect of the decay rate $i \Sigma_0$ is negligible. For more details about the calculation the reader should turn to Appendix \ref{app:selfenergy}.

In the absence of decay, no zero energy states are introduced by the disorder. The effect of $\Sigma_1$ and $\Sigma_3$ is to renormalize the gap and the hopping, leading to a corrected spectrum $\tilde{E}_{\bf k}=[\tilde{\xi}_{\bf k}^2 + \tilde{\D}_{\bf k}^2]^{1/2}$. In the low energy limit this is equivalent to a renormalization of the effective values of $v_f$ and $v_\D$ which we find to be,
\bea\label{vfvd-ren}
\tilde{v}_f &=&v_f\left[1+ 2n_i \frac{\delta_t}{t_{\rm eff}}\left(1-\eta\right) \right]\nn\\
\tilde{v}_\D&=& v_\D \left[1+2n_i \frac{\delta_\D}{\D_0} \left(1-\eta\right)\right].
\eea
The renormalization of $v_\Delta$ is the anti-proximity effect due to the metallic inclusions, which gives rise to a modified coefficient of the linear DOS compared with the pure system. These modifications primarily affect the low temperature physics in the disordered system.

With the Green's function at hand we can calculate the leading contributions to the superfluid stiffness. Details of the calculations appear in Appendices \ref{app:diamagnetic} and \ref{app:paramagnetic}.
The contributions to the superfluid stiffness, to second order in the disorder strength, can be separated into zero temperature and finite temperature contributions.

\emph{  Zero temperature.--} The contribution to the zero temperature stiffness due to self energy corrections is the diamagnetic response expressed in diagram $D_0$. This is a non-universal contribution which turns out to differ only very slightly from the bare diamagnetic stiffness of the pure system (see Appendix \ref{app:diamagnetic} for details),
\bea\label{D0}
D_0 &=& 2gt \sum_{\bf k} \cos{k_x} \left(1-\frac{\tilde{\xi}_{\bf k}}{\sqrt{\tilde{\xi}^2_{\bf k} + \tilde{\Delta}^2_{\bf k}} }\right)\nn\\
&=&gt \mathcal{D}_0\left(\frac{\tilde{v}_\D}{\tilde{v}_f}\right)
\eea
where $\mathcal{D}_0(X)$ is an order unity slowly decreasing function of its argument in the relevant range of parameters. Note that in practice, this function may include a weak dependence on the chemical potential which we neglect, assuming low doping.
To conclude, the renormalization of the spectrum parameters due to the self energy corrections have a negligible effect on the diamagnetic stiffness.\\

\emph{  Finite temperature.--} The finite temperature contribution to the stiffness due to self energy corrections arises from diagram $\Pi_0$. The effect of disorder here is to modify the low energy density of states through a renormalization of the effective values of $v_f$ and $v_\D$.  This affects the superfluid stiffness through the paramagnetic contribution $\Pi_0$ leading to faster reduction of the stiffness with temperature. More precisely
\bea\label{Pi0}
\Pi_0 &=& - \frac{2\log{2}}{\pi }\ \frac{8(Zt)^2}{\tilde{v}_f \tilde{v}_\D}\ T\\
&\simeq&-\left[1-4n_i \left(\frac{\delta_\D}{\D_0} + \frac{\delta_t}{t_{\rm eff}} \right)\left( 1-\eta\right)\right]b_0 T\nn.
\eea
Here $b_0=-d\rho_s/dT$ is the slope in the clean system and $Z$ is the renormalization of the quasiparticle current by interactions. The low $T$ behavior is dominated by low energy quasiparticles, which may be altered by Fermi-liquid renormalization not included in the mean field theory. Therefore $Z$ should be taken as a phenomenological Fermi-liquid parameter\cite{Millis:1998,Ioffe:2002} and not as the value $g(p)$ dictated by the microscopic mean field theory.

In our case the disorder acts to induce faster decrease of the superfluid stiffness with temperature. This is because when the inclusions are highly overdoped with nearly zero gap then $\d_\D/\D_0\gtrsim -1$, while $0<\d_t/t_{\rm eff}\ll1$.\\

%: fig:disorder_rhoTc
\begin{figure}[h]
\begin{center}
\includegraphics{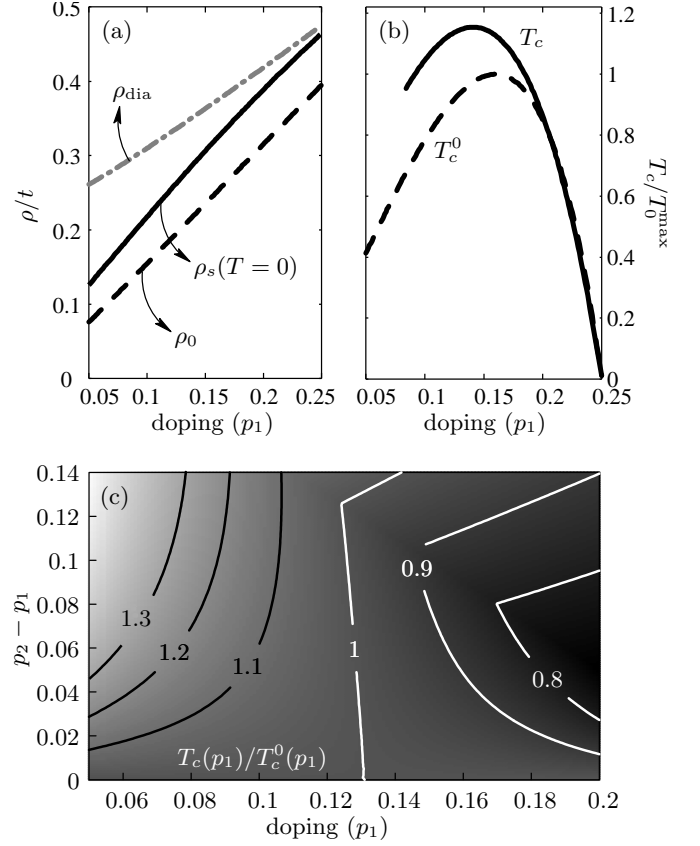}
\end{center}
\caption{\emph{Superfluid stiffness and critical temperature in the inhomogeneously doped layer with point impurities. } In panels (a) and (b) the impurities have an average doping charge $p_2=0.35$ while the bulk doping $p_1$ varies, with impurity concentration $n_i=0.1$. (a) Zero temperature stiffness (Solid) compared to that of the clean case (dashed). The Dashed-dotted line marks the diamagnetic contribution. (b) $T_c$ with impurities compared to $T_c^{0}$ without.  We used a doping independent quasiparticle current renormalization of $Z=0.5$. Results are plotted only within the validity range of the diagrammatic expansion. (c) Relative change in $T_c$ for the case of small excess doping on the impurity, with impurity concentration $n_i=0.2$. Contours map the relative change $T_c/T_c^{0}$ as a function of the base doping $p_1$ and the excess doping on the impurities $p_2-p_1$.}
\label{fig:disorder_rhoTc}
\end{figure}
%%%%%%%%%%%%%%%%%%%%%%%

 {\bf Current operator renormalizations.} The second type of corrections to the stiffness have no counterpart in systems with standard on-site disorder. The disorder in the hopping amplitude introduces renormalizations of the kinetic energy $K_x$  and the current operator $j_x$, proportional to $\delta_t^{c}$. This leads to corrections of $O(\delta_t^{c}), O[(\delta_t^{c})^2]$ to the stiffness, that are represented as vertex corrections in diagrams $D_1,D_2,\Pi_2,\Pi_3,\Pi_4$ of Fig.~\ref{fig:diagrams}. We again distinguish between zero temperature and finite temperature contributions to the stiffness.\\

\emph{  Zero temperature.--}
The most intuitive effect of the vertex correction is the increase of the diamagnetic stiffness at the impurity sites due to the extra charge carriers they contribute. This effect is reflected in the diagram $D_1$ with each impurity bringing an additional $2\delta_t^{c}$ to the average kinetic energy

\bea\label{D1}
D_1 &=& 4n_i\delta_t^{c} \sum_{\bf k} \cos{k_x} \left(1-\frac{{\xi}_{\bf k}}{\sqrt{{\xi}^2_{\bf k} +{\Delta}^2_{\bf k}} }\right)\nn\\
&=&2 n_i\delta_t^{c} \mathcal{D}_0\left(\frac{v_\D}{v_f}\right)\simeq 2 n_i\frac{\delta_t^{c}}{gt} \rho_0.
\eea
This expression reveals a small parameter, $n_i\d_t^{c}/gt $, that did not appear in the Hamiltonian.
The perturbative correction inevitably becomes large upon underdoping towards the Mott insulator where $n_i\d_t^{c}/gt \to \infty$. This signals the breakdown of the Born approximation at doping levels smaller than $p_1^*\simeq g(p_2) n_i/(1+2n_i)$.

The second significant contribution to the zero temperature stiffness stems from the paramagnetic diagram $\Pi_4$, which is seen to be
 \bea\label{Pi4}
\Pi_4 &=& - 2 n_i \frac{(\delta_t^{c})^2}{t_{\rm eff}}\mathcal{P}_0\left(\frac{v_\D}{v_f}\right).
\eea
Here $\mathcal{P}_0(X)$ is an order unity decreasing function of its argument in the relevant parameter range. This term is closely analogous to the zero temperature paramagnetic reduction in the stripe model of sec.~\ref{section:stripes}. Here as in the stripe model, The effect acts to moderate the  enhancement of the stiffness at zero temperature.

Another correction to the zero temperature stiffness is given by the diagram $D_2$. This diagram, which represents a combined renormalization of the vertex and the spectrum, is calculated to be
\bea\label{D2}
D_2 &=&4n_i\delta_t^{c} \left[\frac{\d D}{\D_0} \mathcal{D}_1\left(\frac{v_\D}{v_f}\right) +  \frac{\delta_t}{t_{\rm eff}} \mathcal{D}_2\left(\frac{v_\D}{v_f}\right)\right]\nn.
\eea
Here $\mathcal{D}_1(X)$ is an increasing function and $\mathcal{D}_2(X)$ is slowly decreasing, and their weak dependence on the chemical potential is again neglected.
This diagram turns out to give a negligible numerical contribution to the overall stiffness.

\emph{Finite temperature.--}
The finite temperature contributions to the stiffness that arise from current renormalization are shown in diagrams $\Pi_2$ and $\Pi_3$. An explicit calculation gives
\bea\label{Pi2Pi3}
\Pi_2 &=& - 4 n_i \frac{\delta_t^{c}}{gt}\frac{2\log{2}}{\pi }\ \frac{(2\sqrt{2}Zt)^2}{\tilde{v}_f \tilde{v}_\D}\ T\nn\\
\Pi_3 &=& 4 n_i\  \eta\  \frac{\delta_t^{c}}{gt}\frac{2\log{2}}{\pi }\ \frac{(2\sqrt{2}Zt)^2}{\tilde{v}_f \tilde{v}_\D}\ T.
\eea
Within SBMFT the current renormalization $Z=g(p)$ depends strongly on the doping. However, it is known that this strong doping dependence leads to a disagreement with the experimentally measured slope $d\rho_s/dT$, which is seen to be almost independent of doping.\cite{Wen:1998}

Here we adopt a phenomenological approach, with an effective paramagnetic current renormalization $Z$ which is independent of doping.\cite{Paramekanti:2002, Millis:1998} This holds at finite low temperature, when the physics is dominated by the effective theory of low energy Dirac quasiparticles. In this case, the entire contribution $\Pi_2+\Pi_3$ is negligible because it stems precisely from the difference in the local current operator between the $p_1$ superconductor and the $p_2$ impurities. Therefore, when summing up the finite $T$ contributions to the stiffness we neglect these two diagrams.

\subsection{Results and Discussion}\label{subsec:results2}
We can summarize the results of this section by putting together the various contributions to the superfluid stiffness. This gives the temperature dependent stiffness
\be
\rho_s(T)=\rho_0-b_0 T +2 n_i(\d\rho_s(0)-\d b\  T)
\label{rho-final}
\ee
Here the first two terms constitute the usual expression for the temperature dependent superfluid stiffness of a uniform $d$-wave superconductor\cite{Lee:1997}, as reviewed in sec. \ref{section:overview}.
The second term is
\be
\d\rho_s(0)=\frac{\delta_t^{c}}{gt}\rho_0-\frac{(\delta_t^{c})^2}{t_{\rm eff}}\mathcal{P}_0\left(\frac{v_\D}{v_f}\right)
\ee
The leading order correction to $\d\rho_s(0)$ in the impurity strength is due to the added charge carriers donated by the impurities. The negative
second order term is a paramagnetic correction to the zero temperature stiffness analogous to
the paramagnetic correction that we derived previously for a bilayer heterostructure. In the latter case
this correction was proportional to $(J_1-J_2)^2$, the square of the difference of the quasi-particle currents on the two layers. Here similarly this contribution scales as $(\delta_t^{c})^2\propto[g(p_2)-g(p_1)]^2$,
which is the square of the difference between the local current renormalization in the bulk and near the impurity.

The last term in Eq. \ref{rho-final} is the change of the linear reduction of the stiffness with temperature
due to the presence of impurities. It is given by
\be\label{stiffness:disorder}\nn
\d b=2b_0 \left(-\frac{\delta_\D}{\D_0} - \frac{\delta_t}{t_{\rm eff}} \right)  \left(1-S_1 \frac{\delta_\D}{\D_0} +(1+S_1) \frac{\delta_t}{t_{\rm eff}} \right)
\ee
where $S_1=2\sqrt{v_\D/ v_f}/\pi$ and $b_0$ is the parameter for the uniform superconductor given in sec. \ref{section:overview} $b_0=2\log{2} (2\sqrt{2}Zt)^2/(\pi v_f v_\D)$\cite{Lee:1997}. Note that the expression in the first bracket is positive because $\d\D <0$ on the impurities. Hence the superfluid stiffness is reduced faster as a function of temperature than in the uniform superconductor. We estimate the parameters of the uniform system using SBMFT, so that $v_f=2\sqrt{2}t_{\rm eff}$ and $v_\D=\sqrt{2}\D$. Taking $J=t/3$, the effective hopping and gap parameters are given by $t_{\rm eff}(p)=g(p) t+\chi$ and $\D(p)=\chi[1-4 p]$, where $\chi$ is the value of the mean fields (both pairing and Fock field) at zero doping.

Figure~\ref{fig:disorder_rhoTc}(a) displays the calculated zero temperature stiffness as function of the doping $p_1$ with the impurities fixed to a high doping level $p_2=0.35$, which corresponds to zero pairing amplitude, and a hopping amplitude of $t$. We plot the total stiffness $\rho_s(T=0)$ as well as the diamagnetic contribution $\rho_{\rm dia}=D_0+D_1+D_2$. Note that the diamagnetic contribution in the disordered system $\rho_{\rm dia}$ is significantly increased with respect to the pure case, $\rho_0$. However, the total zero temperature stiffness $\rho_s(T=0)$ is only moderately increased compared to the uniform case (where $\rho_0$ is the total stiffness at $T=0$). The reason for this is the zero temperature paramagnetic contribution of the impurities $\Pi_4$.

In panel (b) of the same figure we plot the critical temperature as a function of the bulk doping $p_1$, estimated from Eq. (\ref{stiffness:disorder}) using the Kosterlitz-Thouless criterion $\rho_s(T_c)=2T_c/\pi$. Again this is for a fixed value of $p_2=0.35$ and $n_i=0.1$ and the result is compared against $T_{c}^0$ of the pure system. The critical temperature of the disordered system is significantly enhanced, above the maximal $T_c$ of the clean superconductor. The maximum of $T_c$ is shifted to the underdoped regime. These results are reminiscent of experiments by Yuli et al\cite{Yuli:2008} that show a $T_c$ enhancement in an underdoped-overdoped bilayer. We can relate our results to the experiment if we assume that the interface between the two layers is in fact an inhomogeneous mixture of underdoped and overdoped regions. Our results suggest that an optimal configuration for $T_c$ enhancement can be achieved by placing point-like metallic inclusions inside a slightly underdoped superconductor.

 Figure \ref{fig:disorder_rhoTc}(c) shows the relative change in the critical temperature with respect to $T_c^{0}$ as function of $p_1$ and $\d p=p_2-p_1$, for $n_i=0.2$. The critical temperature is enhanced relative to the clean system by up to $15\%$, in a broad range of $p_1$ and $\d p$. Here the excess doping on the impurities $\d p$ is small, and there is no enhancement of $T_c$ above the maximal $T_c$ of the clean superconductor. The main reason for this is the zero temperature paramagnetic reduction of the stiffness due to the impurities. Without this effect we could have obtained an absolute enhancement of $T_c$ in the disordered system, also in the small $\d p$ limit. We have checked and found that whether we use the microscopic or phenomenological parameter $Z$ to renormalize the quasiparticle current makes very little difference to the final result of $T_c$.\\

It is instructive to look at the behavior of the stiffness and $T_c$, for small values of $\d p\equiv p_2-p_1$, for which we can neglect second order contributions in $\delta_t/t_{\rm eff}$ and $|\delta_\D/\D_0|$, such as the paramagnetic effect. Here we use the SBMFT doping dependence for both the bulk and the impurity, such that $\delta_t=\delta_t^{c}=t[g(p_2)-g(p_1)]$. In this regime there is a simple expression for the superfluid stiffness,
\be
\rho_s(T)\simeq \rho_{\rm cl}(T)+ 2n_i\frac{\delta_t^{c}}{gt}\rho_0-4n_i \left(\left|\frac{\delta_\D}{\D_0} \right|- \frac{\delta_t}{t_{\rm eff}}\right)b_0 T\nn
\ee
where $\rho_{\rm cl}(T)=\rho_0-b_0T$ is the stiffness of the clean superconductor. The zero temperature stiffness is always enhanced, whereas the slope $|d\rho_s/dT|$ is increased. The latter is easily seen by expressing $|\delta_\D/\D_0|-\delta_t/t_{\rm eff}$ as function of $p_1$ and $\d p$.
Using the Kosterlitz-Thouless criterion as above we can estimate the change in transition temperature $T_c$ with respect to the critical temperature $T_c^{0}$ of the clean superconductor,
\be
\frac{T_c}{T_c^{0}}\simeq1+ 2n_i \left[\frac{\delta_t^{c}}{gt}+\frac{2b_0}{b_0+2/\pi} \left(\frac{\delta_t}{t_{\rm eff}} -\left|\frac{\delta_\D}{\D_0} \right|\right) \right].
\ee
This can be expressed in terms of the doping level $p_1$ of the clean superconductor and the difference in doping $\d p=p_2-p_1$ between the background and the impurities. We obtain an expression of the form
\be
\nn
\frac{T_c}{T_c^{0}}
=1+ 2n_i
\frac{\d p}{p_1}(p^*-p_1)\mathcal{F}(p_1).
\label{Tcdp}
\ee
where $\mathcal{F}(p_1)$ is positive for $p_1<0.25$. This implies that for $p_1<p^*$, the critical temperature of the disordered system is enhanced compared with the clean superconductor with doping $p_1$.
Under the assumptions of SBMFT, with $Z=0.5$ and $\chi=0.4t$, we get $p^*\sim0.125$ and
\be
\mathcal{F}(p_1)\simeq\frac{(1\!-\!p_1)(0.08\!+\!p_1)(0.6+\!p_1)}{(0.5\!-\!p_1)(0.25\!-\!p_1)(0.17\!+\!p_1)(0.29+p_1)}.
\ee
We note that $\mathcal{F}(0.1)\simeq 18$.

It seems from Eq. (\ref{Tcdp}) that $T_c$ can be further enhanced by increasing the impurity concentration. However, by doing
this we would quickly violate the Born approximation. In particular, the superfluid stiffness in this non perturbative regime should be calculated as the resistance of an effective resistor network with $\rho_s^{-1}$ of the various puddles playing the role of the resistance.\\

We now remark on the nature of our perturbative approach and the small parameters involved in it. The scattering from individual impurities is taken into account within the Born approximation to second order in the impurity strength as measured by the parameters $\delta_t/t_{\rm eff}$ and $|\delta_\D/\D_0|$. We found that the second order correction to both $\rho_s$ and $T_c$ was always negligible compared to the first order contribution. This was the case even when we took the parameter $|\d \D/\D|\simeq 1$.
An additional small parameter $n_i\d_t^c/gt$ appeared through the effect of the impurities on the coupling to the electromagnetic field rather than the scattering on the impurity potential.

We would like to contrast our approach with the commonly used self consistent $T$-matrix approximation (SCTMA)\cite{comment-SCTMA}, which treats the single impurities exactly. This turns out to be important to describe the effect of in-plane ion substitutions such as Zn impurities that act as unitary scatterers and give rise to strong bound states. However in our case the SCTMA is not analytically solvable because of the strong momentum dependence of the impurity potential and the fact that it acts as a matrix in Nambu space ($\d_t$ is the diagonal component and $\d_\D$ off-diagonal). Fortunately, the disorder potential that interests us is generated by dopants, which reside outside the CuO plane.\cite{comment-Yttrium} \nocite{MacKenzie:1994} Indeed we can show that the fact that such impurities induces only a small change in the local doping level ($p_2-p_1\ll 1$) implies that the impurity scattering is far from the unitary limit and does not give rise to a bound state. To see this consider
the strength of the local impurity potential $U_0\simeq (p2-p1)/\kappa$. Since the compressibility $\kappa$ is approximately the density of states at the Fermi level $\nu_0$, the dimensionless impurity strength is just $U_0\nu_0\simeq p_2-p1\ll 1$. This is far from satisfying the condition for formation of a bound state.\cite{Balatsky:1995} In this limit the direct impurity scattering can be taken within the Born approximation and lead to negligible contributions to the low energy DOS.\cite{Durst:2000,Sharapov:2002} Hence it leads to a concomitantly small correction to the stiffness. 

\section{Conclusions}\label{section:summary}
%%%%%%%%%%%%%%%%%%%%%%%
We investigated the effects of doping inhomogeneity on the superconducting properties of the cuprates using the slave boson mean field theory\cite{Kotliar:1988} supplemented by phenomenological Fermi liquid parameters to account for the low energy quasiparticle properties\cite{Ioffe:2002}. In particular the superfluid stiffness and the critical temperature was calculated within two different models of the inhomogeneity.

The first model described doping variations on mesoscopic scales, comparable to or larger than the superconducting coherence length. Technically we computed the transverse electromagnetic response tensor of a model system with metallic stripes embedded in an underdoped superconducting bulk. This was done by solving the appropriate Bogoliubov-de Gennes equations within the renormalized mean field theory. We then averaged over the different stripe orientations to obtain an isotropic superfluid stiffness.

In the second model we considered microscopic impurities that carried an excess doping charge.
The temperature dependent stiffness in this case was calculated using a perturbative expansion  expansion assuming both dilute and weak impurities (Born approximation).

In both models, the regions of higher doping add to the total carrier density and hence increase the
superfluid stiffness at zero temperature. On the other hand the impurity regions give rise to low energy states that lead to a faster reduction of the superfluid stiffness with temperature. Nevertheless we found that for a range of doping levels in the underdoped regime a higher $T_c$ than a uniform superconductor of the same doping can be attained. Moreover, in the case of microscopic impurities it is even possible to attain a higher critical temperature than the maximal $T_c$ obtained in the pure system, that is, higher than $T_c$ at optimal doping.

The last result can help to understand the enhancement of $T_c$ seen at the interface between an underdoped and a highly overdoped LSCO film.\cite{Yuli:2008} We have previously noted that such an increase in $T_c$ due to coupling between two homogeneous layers with different doping requires unrealistically strong coupling between the two CuO planes.\cite{Goren:2009} However if, due to the structure of the interface, overdoped and underdoped layers interpenetrate each other, then the proximity coupling can be induced by the strong in-plane hopping and the situation becomes equivalent to the one considered here.

Finally we remark that our main result for the case of microscopic impurities was obtained within a perturbative expansion in the impurity strength. It would be interesting to compare this to a numerical solution that takes into account scattering, at least from individual impurities, exactly.
If indeed excess dopants concentrated at random locations can lead to increase of the maximal $T_c$, this opens up intriguing possibilities for further enhancement of $T_c$. For example through design of an optimal ordered arrangement of the highly doped regions.

\section{Acknowledgments}
%%%%%%%%%%%%%%%%%%%%%%%
We thank H. Bary-Soroker, E. Berg, E. Demler, S. Huber, Y. Kraus, K. Michaeli, and Z. Ringel for valuable discussions. This work was supported by grants from the Israeli Science Foundation and the Minerva foundation.

\appendix

\section{The self energy in Born approximation}\label{app:selfenergy}
%%%%%%%%%%%
We write down the Dyson equation for the disorder averaged Green's function \cite{comment-Doniach},
\be
G_{{\bf k},\omega_n}=G_{{\bf k},\omega_n}^0 + G_{{\bf k},\omega_n}^0 \Sigma({\bf k},\omega_n)G_{{\bf k},\omega_n}.
\ee
From  the Dyson equation we obtain the disorder averaged self energy, up to second order in  the disorder potential $\mathcal{U_{\bf kk'}}\equiv{U}_{\bf kk'}\s_3  + {V}_{\bf kk'}\s_1$,
\be\label{selfenergy:Born}
\Sigma({\bf k},\omega_n)=n_i[\mathcal{U_{\bf kk}}+ \sum_{\bf k'}\mathcal{U_{\bf kk'}} G_{{\bf k'},\omega_n}\mathcal{U_{\bf k'k}}].
\ee
Using (\ref{selfenergy:Born}) we can now calculate the Nambu components of $\Sigma({\bf k},\omega_n)=\sum_{a=0}^3 \s_a \Sigma_a$.
\bea
\Sigma_0({\bf k},\omega_n)=-i\omega_n n_i \sum_{\bf k'} \frac{U_{\bf kk'}^2+V_{\bf kk'}^2}{\omega_n^2+E_{\bf k'}} \nn
\eea
with $E_{\bf k}=[\xi_{\bf k}^2+\D_{\bf k}^2]^{1/2}$.
In the limit of small $\omega_n$ this becomes
\begin{multline}\label{Sigma0}
\Sigma_0({\bf k},\omega_n)
\simeq  -i\omega_n n_i \left[\left( \frac{\delta_\D}{\D_0}\right)^2 S_1 + \left( \frac{\delta_t}{t_{\rm eff}}\right)^2 (1-S_1)\right]\\
-i\omega_n n_i  \left[\left( \frac{\delta_\D}{\D_0}\right)^2 \D_{\bf k}^2 +\left( \frac{\delta_t}{t_{\rm eff}}\right)^2 (\xi_{\bf k}+2\mu)^2 \right] S_0(\omega_n)  
\end{multline}
where $S_1=\lim_{\omega_n\ra 0}\sum_{\bf k} \D_{\bf k}^2/(\omega_n^2 + E_{\bf k}^2)\simeq \sqrt{2\D_0/t_{\rm eff}}/\pi$. In (\ref{Sigma0}) we used the fact that in the limit of $\omega_n\ra 0$, $S_1=\sum_{\bf k} \D_{\bf k}^2/(\omega_n^2 + E_{\bf k}^2)\simeq 1-\sum_{\bf k} \xi_{\bf k}^2/(\omega_n^2 + E_{\bf k}^2)$.
The sum $S_0$ is logarithmically divergent in the $\omega_n\ra 0$ limit,
\be
S_0=\sum_{\bf k} \frac{1}{\omega_n^2 + E_{\bf k}^2}=\frac{1}{4\pi v_f v_\D}\log{\left[1-\frac{C^2}{(i\omega_n)^2}\right]}
\ee
Where $C$ is an upper cutoff for the momentum sum. To solve for the zero frequency limit of the self energy we follow Ref.~\onlinecite{Lee:1993} and assume a self consistent solution of the form $\Sigma_0({\bf k},\omega\ra 0)\ra -i \G_{\bf k}$. For the self consistent solution we perform the analytic continuation $i\omega_n\ra \omega+i\d$ and replace $\omega$ by its renormalized value $\tilde{\omega}=\omega-\Sigma\ra i\G_{\bf k}$. This gives the following equation for $\G_{\bf k}$,
\be
\frac{1}{n_i}= \frac{\bar{U}_{\bf k}^2}{4\pi v_f v_\D}\log{\frac{C^2}{\G_{\bf k}^2}}+\left( \frac{\delta_\D}{\D_0}\right)^2 \!S_1 + \left( \frac{\delta_t}{t_{\rm eff}}\right)^2\! (1-S_1)\nn
\ee
where we denote $\bar{U}_{\bf k}^2\equiv \left( \frac{\delta_\D}{\D_0}\right)^2 \D_{\bf k}^2 +\left( \frac{\delta_t}{t_{\rm eff}}\right)^2 (\xi_{\bf k}+2\mu)^2$. In the limit ${\bf k}\ra {\bf k}_{\rm node}$ we approximate $\bar{U}_{\bf k}^2\sim \mu^2\d_t^2/t_{\rm eff}^2$. Finally, using the fact that $|\left( {\delta_\D}/{\D_0}\right)^2 \!S_1 + \left( {\delta_t}/{t_{\rm eff}}\right)^2\! (1-S_1)| \le 1 \ll 1/n_i$ we obtain the low frequency long wavelength limit of the decay rate $\G$,
\be
\G=C e^{-\frac{v_f v_\D t_{\rm eff}^2}{n_i(2\mu\delta_t)^2}}.
\ee
Where $C\simeq\D_0$ is a high energy cutoff. We shall now estimate the exponent and show that $\G$ is negligible. We plug in the doping dependent values
$\mu\propto p_1t , t_{\rm eff}/\delta_t=v_f/[2\sqrt{2}(g(p_2)-g(p_1))t], v_\D/v_f \sim 0.5(1-4 p_1) $ and $n_i  <0.25$. We obtain $\G \lesssim \D_0 e^{-100}\simeq0$.

Using the fact that $\Sigma_0\ra 0$, we can calculate the other two components of the self energy,
\bea
\Sigma_1({\bf k},\omega_n)&=&n_i V_{\bf kk}  + \nn\\
&+&n_i \sum_{\bf k'} \frac{\D_{\bf k'}(U_{\bf kk'}^2-V_{\bf kk'}^2)-2\xi_{\bf k'}U_{\bf kk'}V_{\bf kk'}}{\omega_n^2+E_{\bf k'}} \nn\\
\Sigma_3({\bf k},\omega_n)&=&n_i U_{\bf kk}  + \nn\\
&+&n_i \sum_{\bf k'} \frac{\xi_{\bf k'}(V_{\bf kk'}^2-U_{\bf kk'}^2)-2\D_{\bf k'}U_{\bf kk'}V_{\bf kk'}}{\omega_n^2+E_{\bf k'}} \nn.
\eea
We perform the momentum summations in the $\omega_n \ra 0$, and express the results in terms of $S_1$ as in the case of $\Sigma_0$. This gives Eqns.~(\ref{selfenergy}).

\section{Diamagnetic response}\label{app:diamagnetic}
The diamagnetic response stems from the second order term in the vectors potential,
\bea
H_{\rm dia}&=&-\frac{1}{2}\sum_{\bf r} K_x({\bf r}) A_x^2({\bf r})\\
&=&-\frac{1}{2}\sum_{\bf q,q'} A_x({\bf q}) K_x({\bf -q-q'}) A_x({\bf q'})\nn
\eea
where $k_x({\bf r})=-\sum_{{\bf r},\s} t_x({\bf r})(c_{{\bf r}+x,\s}^\dag c_{{\bf r},\s}+h.c$. Its Fourier transform to momentum space is then
with
\bea
&K_x&({\bf -q-q'}) \equiv \sum_{\bf r} e^{i({\bf q+q'} ){\bf r}} K_x({\bf r}) \\
&=&\sum_{\bf r,k,k',\s} t({\bf r}) e^{i({\bf q+q'+k-k'} ){\bf r}}\left(e^{ik_x}+e^{-ik'_x}\right) c_{{\bf k},\s}^\dag c_{{\bf k'},\s}.\nn
\eea
Performing the disorder average leads to a diagrammatic expansion with three contributions,
\bea
\overline{\langle K_x({\bf -q-q'})\rangle} &=&D_0  + D_1 +D_2  \nn
\eea
where $D_0$ is the diamagnetic contribution including only self-energy corrections to the Green's function, described by diagram \ref{fig:diagrams}(a). An explicit calculation of this diagram gives
\be
D_0 =g t\sum_{\bf k,k'} \delta_{\bf q+q'+k-k'} \left(e^{ik_x}+e^{-ik'_x}\right) \tr\{{G_{\bf k}}\s_3\}\delta_{\bf kk'}\nn.
\ee
Note that in our notations the trace includes the Matsubara summation and Nambu space tracing. The contribution $D_1+D_2$ of diagrams \ref{fig:diagrams}(b) and (c) is due to modification of the hopping on the impurity sites and is therefore proportional to $\delta_t^{c}$,
\be
D_1+D_2= \delta_t^{c} \sum_{\bf k,k'}\left(e^{ik_x}+e^{-ik'_x}\right)  \tr\{ \overline{\rho_{\bf q+q'+k-k'} G_{\bf kk'}} \s_3\}\nn
\ee
 The object $G_{\bf kk'}$ is defined by
 \bea\label{Gkk'}
G_{\bf kk'}&\equiv& G_{\bf k}^0\delta_{\bf kk'}+ G_{\bf k}^0 \hat{\rho}_{\bf k'-k} \mathcal{U_{\bf kk'}}G_{\bf k'}^0 + \nn\\
&+&\sum_{\bf p} G_{\bf k}^0 \hat{\rho}_{\bf p-k} \mathcal{U_{\bf kp}}G_{\bf p}^0 \rho_{\bf k'-p} \mathcal{U_{\bf pk'}}G_{\bf k'}^0 \nn
\eea
As usual, the average over realizations amounts to integrating over all possible impurity positions. In all the summations, the only dependence on impurity positions appears in factors of $\hat{\rho}_{\bf k}$. The disorder averaging gives
\bea
\overline{\hat{\rho}_{\bf k}}&=&n_i \int d^3 {\bf r}e^{-i{\bf k r}}=n_i \delta_{\bf k}\\
\overline{\hat{\rho}_{\bf k}\hat{\rho}_{\bf k'}}&=&\sum_{ij}\overline{e^{-i{\bf k r_i}}  e^{-i{\bf k' r_j}}   }
\simeq n_i \delta_{\bf k+k'} + O(n_i^2)\nn
\eea
We perform the sums and and keep terms up to first order in $n_i$ and second order in the disorder strength $\d_t$ and $\d_\D$. This gives
\bea
D_0 &=& \delta_{\bf q+q'} 2gt \sum_{\bf k} \cos{k_x} \ n(\tilde{\xi}_{\bf k},\tilde{\Delta}_{\bf k})\\
D_1&=&\delta_{\bf q+q'} 4 n_i\delta_t^{c} \sum_{\bf k} \cos{k_x} \ n(\xi_{\bf k},\Delta_{\bf k}) \nn\\
D_2&=&\delta_{\bf q+q'} 2 n_i\delta_t^{c} \sum_{\bf kk'} \left(e^{ik_x}+e^{-ik'_x}\right)  \tr\{ G_{\bf k}^0 \mathcal{U_{\bf kk'}}G_{\bf k'}^0 \s_3\}\nn
\eea
where $n(\xi,\Delta) \equiv [1-\xi/(\sqrt{\xi^2+\D^2})]$. Note that $\xi$ and $\D$ appear in their renormalized values in $D_0$ and are unrenormalized in $D_1$.

Since $D_2$ is non vanishing at $T=0$, it does not depend necessarily on low energy quasiparticles. Indeed, it includes a sum over all occupied states. We evaluate it numerically and express the result as a function of $\nu=\D_0/2t_{\rm eff} = v_\D/v_f$ in the limit of half filling. Any deviation from half filling introduces a small value of the chemical potential $\mu$ which we neglect in this calculation. The result has the general form
\be
D_{2}=4n_i\delta_t^{c} \left[\frac{\d D}{\D_0} \mathcal{D}_1(\nu) +  \frac{\delta_t}{t_{\rm eff}} \mathcal{D}_2(\nu)\right]
\ee
and turns out to give a negligible numerical contribution in the relevant regime of parameters.

\section{Paramagnetic response}\label{app:paramagnetic}
%%%%%%%%%%
the paramagnetic current-current correlator for a given disorder realization is
\be\label{paramagnetic:definition}
\Pi_{xx}(q,q',\omega)=-\int_0^\b d\tau e^{i\omega_n \tau} \langle j_x(q,\tau) j_x(-q', 0)\rangle.
\ee
Naturally, after disorder averaging all contributions are proportional to $\delta_{\bf qq'}$.
The current operator is modified by the disordered hopping, and has the form $j({\bf q})= j^{(0)}({\bf q})+\d j({\bf q}) $. The uniform part of the current operator in the $x$ direction is
\bea
j_x^{(0)}({\bf q})& =& igt\sum_{{\bf k}\s}(e^{ik_x} - e^{-i(k_x-q_x)}) c_{{\bf k}\s}^\dag c_{{\bf k-q},\s}\nn\\
&\equiv& J_{\bf kq}c_{{\bf k}\s}^\dag c_{{\bf k-q},\s}\nn.
\eea
The disorder contribution to the current operator is given by
\be
\d j({\bf q}) = i\d_t^{c} \sum_{{\bf kk'}\s} \rho_{\bf q-k+k'}(e^{ik_x} - e^{-ik'_x})c_{{\bf k}\s}^\dag c_{{\bf k'}\s}.
\ee
As a result, the current-current correlator has the form
\begin{multline}\label{jjcorr}
\langle j_x({\bf q},\tau)  j_x({\bf -q'}, 0)\rangle= \langle j^{(0)}_x({\bf q},\tau) j^{(0)}_x({\bf -q'}, 0)\rangle\\
 +\frac{2\delta_t^{c} }{gt}\sum_{\bf k}\hat{\rho}_{\bf q-k} \langle j_x^{(0)}({\bf k},\tau) j_x^{(0)}({\bf -q'}, 0)\rangle \\
 + \left(\frac{\delta_t^{c} }{gt}\right)^2 \sum_{kk'} \hat{\rho}_{\bf q-k}\hat{\rho}_{\bf k'-q'} \langle j_x^{(0)}({\bf k},\tau) j_x^{(0)}({\bf -k'}, 0)\rangle .
\end{multline}

The disorder averaging of $\Pi_{xx}(q,q',\omega)$ amounts to averaging the correlator (\ref{jjcorr}) over realizations.

The first line of (\ref{jjcorr}) corresponds to diagrams $\Pi_0+\Pi_1$ in Fig.~\ref{fig:diagrams}, which incorporate the effects of self energy renormalization and standard vertex corrections,
\bea
\Pi_0+\Pi_1&=&-\int_0^\b d\tau e^{i\omega_n \tau}\overline{ \langle j^{(0)}_x(q,\tau) j^{(0)}_x(-q', 0)\rangle}\nn.
\eea
The disorder averaged correlator takes the form
\begin{multline}
\overline{ \langle j^{(0)}_x(q,\tau) j^{(0)}_x(-q', 0)\rangle}=\\
 \delta_{\bf qq'} \sum_{\bf kk'} J_{\bf kq}J_{\bf k',-q}\tr\{ \overline{ G_{\bf k'+q,k}(-\tau) G_{\bf k-q,k'}(\tau) }\}.
\end{multline}
The disorder averaged Green's function product has a vertex correction part $\Pi_1$ which vanishes, as we will show below. As a result we are left with a simple product of disorder averaged Green's function,
\bea
\Pi_0&=&-\int_0^\b d\tau e^{i\omega_n \tau}\delta_{\bf qq'}\\
&&\times \sum_{k} |J_{\bf kq}|^2\tr\{ \overline{G_{\bf k'+q,k}(-\tau)}\ \  \overline{G_{\bf k-q,k'}(\tau)}\}\nn\\
& =&-\int_0^\b d\tau e^{i\omega_n \tau}\delta_{\bf qq'}  \sum_{k} |J_{kq}|^2\tr\{ G_{\bf k}(-\tau){G_{\bf k-q}}(\tau)\}\nn.
\eea
To calculate $\Pi_0$ we notice that the disorder averaged Green's function $G_{{\bf k},\omega_n}$ differ from the bare one $G^{0}_{{\bf k},\omega_n}$ by renormalized values of $\D_{\bf k}$ and $\xi_{\bf k}$, as specified in (\ref{selfenergy}), leading to renormalized spectrum parameters $v_f$ and $v_\D$ according to (\ref{vfvd-ren}). This gives the result shown in (\ref{Pi0}).

In order to see that the standard vertex correction $\Pi_1$ vanishes, we write it explicitly as
\begin{multline}
\Pi_1\!=\! n_i \int_0^\b d\tau e^{i\omega_n \tau}\sum_{\bf kk'} J_{\bf kq}J_{\bf k',-q}\\
\!\times\!\tr \{G_{\bf k'+q}^0 \mathcal{U_{\bf k'+q,k}}G_{\bf k}^0 G_{\bf k-q}^0 \mathcal{U_{\bf k-q,k'}}G_{\bf k'}^0\}
\end{multline}
In the limit of ${\bf q}\ra 0$, the sum over momenta becomes $4n_it^2 \sum_{\bf k,k'}\sin{k_x}\sin{k'_x} \mathcal{F}({\bf k,k'})$, where $ \mathcal{F}$ is symmetric with respect to ${\bf k}$ and ${\bf k'}$. Thus, under the summation over ${\bf k}$ or ${\bf k'}$, this contribution vanishes.

 The second line of (\ref{jjcorr}) corresponds to diagrams $\Pi_2+\Pi_3$ and the third line to $\Pi_4$.
These contributions do not appear in the case of standard on-site disorder because they stem from direct renormalization of the current operator $j_x({\bf q})$ by an amount proportional to $\delta_t^{c}$. The first part of this contribution is
\begin{multline}
\Pi_2+\Pi_3
=-\int_0^\b d\tau e^{i\omega_n \tau}\frac{2\delta_t^{c} }{gt}\sum_{kk'p}J_{kp}J_{k',-q'}\nn\\
 \times \overline{\rho_{\bf q-p} \tr\{ G_{\bf k'+q',k}(-\tau)G_{\bf k-p,k'}(\tau) \}}\nn.
\end{multline}

When we insert (\ref{Gkk'}) and perform the disorder average we obtain a contribution of $O(\mathcal{U})$,
\bea
\Pi_2({\bf q},0)&=&-4n_i\frac{\delta_t^{c} }{gt} \frac{1}{\b}\sum_{{\bf k}, n} |J_{\bf kq}|^2\tr\{ G^0_{{\bf k},\omega_n} G^0_{{\bf k-q},\omega_n}\}\nn
\eea
and a contribution of $O(\mathcal{U}^2)$,
\bea
\Pi_3({\bf q},0)&=&-4n_i\frac{\delta_t^{c} }{gt} \frac{1}{\b}\sum_{{\bf kp}, n} J_{\bf kp} J_{\bf k-q,-q} \nn\\
&&\times\ \tr\{ G^0_{{\bf k},\omega_n} G^0_{{\bf k-p},\omega_n} \mathcal{U}_{\bf k-p,k-q} G^0_{{\bf k-q},\omega_n}\}\nn.
\eea
Taking the trace and performing the summations at the limit ${\bf q}\ra 0$ we obtain the results in Eqn.~(\ref{Pi2Pi3}).

Finally, the third line of $(\ref{jjcorr})$, corresponding to second order corrections of the current-current correlation, yields the sum
\bea
\Pi_4({\bf q},0)
&=&-\left(\frac{\delta_t^{c} }{gt}\right)^2  \frac{1}{\b}\sum_{{\bf pp'kk'},n}J_{\bf kp}J_{\bf k',-p'} \nn\\
& &\times\  \overline{ \rho_{\bf q-p}\rho_{\bf p'-q'}  \tr\{ G_{{\bf k'+p',k},\omega_n}G_{{\bf k-p,k'},\omega_n} \} }\nn\\
&=&-\left(\frac{\delta_t^{c} }{gt}\right)^2 \frac{1}{\b} \sum_{{\bf pp'kk'},n}|J_{\bf kp}|^2 \tr\{ G^0_{{\bf k},\omega_n} G^0_{{\bf k-p},\omega_n} \}\nn.
\eea
After summation this gives Eqn.~(\ref{Pi4}).

%\bibliographystyle{phd-url-notitle}
%\bibliography{ref,comments}

\end{document}